\documentclass[11pt]{article}
\usepackage{bbm}

\usepackage{mathrsfs}
\usepackage{amsfonts}
\usepackage{amssymb}
\usepackage{amsfonts, amssymb,  mathrsfs,  amsmath,    amssymb,   theorem,   float, xypic}

\newtheorem{theorem}{Theorem}[section]

\newtheorem{cor}[theorem]{Corollary}
\newtheorem{lemma}[theorem]{Lemma}

\newtheorem{example}[theorem]{Example}

\floatstyle{ruled}
\newfloat{algorithm}{t}{loa}
\floatname{algorithm}{Algorithm}
\newcommand{\Inp}[1]
  {\noindent\begin{tabular}{@{}p{1.8cm}@{}p{13.2cm}@{}}
   {\bf Input: }&#1 \end{tabular}}
\newcommand{\Outp}[1]
  {\noindent\begin{tabular}{@{}p{1.8cm}@{}p{13.2cm}@{}}
   {\bf Output: }&#1 \end{tabular}}

\def\and{\cap}

\def\bref#1{(\ref{#1})}
\def\proof{{\noindent\em Proof:} }

\newcommand{\SPC}{\hspace*{15pt}}
\newcommand{\qedd}{\hspace*{\fill}$\Box$\medskip}

\floatstyle{ruled}
\newfloat{algorithm}{t}{loa}
\floatname{algorithm}{Algorithm}

\def\U{{\mathbb{U}}}
\def\V{{\mathbb{V}}}

\def\A{{\mathcal A}}

\def\D{{\mathcal D}}

\def\T{{\mathcal{T}}}
\def\G{{\mathcal{G}}}

\def\PS{{\mathbb P}}

\def\D{{\mathbb D}}

\def\P{{\mathbb P}}

\def\Zero{\hbox{\rm Zero}}

\def\zero{\hbox{\rm Zero}}

\def\sat{\hbox{\rm{sat}}}

\def\dsat{\hbox{\rm{dsat}}}
\def\vars{\hbox{\rm{vars}}}

\def\deg{\hbox{\rm{deg}}}
\def\cls{\hbox{\rm{cls}}}

\def\ord{\hbox{\rm{ord}}}

\def\ldeg{\hbox{\rm{ldeg}}}
\def\dim{\hbox{\rm{dim}}}

\def\lv{\hbox{\rm{lv}}}

\def\ord{\hbox{\rm{ord}}}

\def\coeff{\hbox{\rm{coeff}}}

\def\ini{\hbox{\rm{ini}}}
\def\rank{\hbox{\rm{rank}}}
\def\ld{\hbox{\rm{ld}}}

\def\Z{{\mathbb{Z}}}

\def\N{{\mathbb{N}}}

\def\R{{\mathbb R}}

\def\F{{\mathbb{F}}}
\def\E{{\mathbb{E}}}

\def\K{{\mathbb{K}}}

\def\and{\cap}

\newcounter{bean}
\def\bl{\begin{list}{Step \arabic{bean}}{\usecounter{bean}}\labelwidth=34pt}
\def\el{\end{list}}

\def\deg{{\rm deg}}

\def\normalization1{{\rm normalization1}}
\def\normalization{{\rm normalization}}

\def\irrfactor1{{\rm irrfactor1}}
\def\irrfactor{{\rm irrfactor}}

\def\zero{{\rm Zero}}
\def\sat{{\rm sat}}

\begin{document}
\title{A Triangular Decomposition Algorithm for Differential Polynomial Systems with Elementary Computation Complexity}
\author{Wei Zhu and Xiao-Shan Gao\\
 KLMM,  Academy of Mathematics and Systems Science\\
 Chinese Academy of Sciences, Beijing 100190, China}
\date{}
\maketitle

\begin{abstract}
\noindent
In this paper, a new triangular decomposition algorithm is proposed for ordinary differential polynomial systems, which has triple exponential computational complexity. The key idea is to eliminate one algebraic variable from a set of polynomials in one step using the theory of multivariate resultant. This seems to be the first differential triangular decomposition algorithm with elementary computation complexity.

\vskip10pt
\noindent
{\bf Keywords}.  Triangular decomposition, regular triangular set, saturated triangular set, differential polynomial system.

\end{abstract}

\section{Introduction}
A basic problem in computer algebra is to properly represent the solutions for a set of algebraic or differential equations and
the triangular set is one of the basic ways to do that.
Let $f_1\ldots,f_s$ be polynomials in variables $x_1,\ldots,x_n$. Then it is possible to compute triangular sets $\T_1,\ldots,\T_r$ such that
 $$\zero(f_1\ldots,f_s) = \cup_{i=1}^r \zero(\sat(\T_i))$$
where $\sat(\T_i)$ is the saturation ideal to be defined in section 2 of this paper. Since each $\T_i$ is in triangular form, many properties of its solution set can be easily deduced. Triangular decompositions also lead to many important applications such as automated theorem proving, kinematic analysis of robotics, computer vision, stability analysis of molecular systems, etc.

The concept of triangular set was introduced by Ritt \cite{ritt1} in the 1950s and was revived in the 1980s by Wu \cite{wu3} in his work of automated geometry theorem proving. One of the major advantage of the triangular decomposition method is that it can be used to give complete methods for the radical ideal membership problem of differential and difference polynomial ideals, while the well known Gr\"obner basis method does not suit for this purpose.
By now, various kinds of triangular decomposition algorithms have been proposed for polynomial systems \cite{lazard,sz1,dmwang1,wu3}, differential polynomial systems \cite{boul,rody1,cgao1,hubert2,wu2}, difference polynomial systems \cite{cs-dd}, polynomial systems over finite fields \cite{cs-f2,huangz1,xli1}, and semi-algebraic sets \cite{chencb1}.

The computational complexity analysis for triangular decomposition algorithms is quite difficult and only very limited results are known.
For polynomial systems, Gallo and Mishra gave a single exponential algorithm to compute the characteristic set for a finitely generated ideal \cite{gm1} and Szanto gave a randomized single exponential algorithm to compute the triangular decomposition \cite{sz1}.
The complexity analysis of the commonly used triangular decomposition algorithms is not given yet. However, it is shown that if solutions in $\Z_2$ are considered, then the commonly used triangular decomposition algorithm can be made single exponential and practically very efficient \cite{cs-f2}.
For differential polynomial systems, it is generally believed that the commonly used triangular decomposition algorithms have non-elementary computational complexity \cite{golu1}.

In this paper, new triangular decomposition algorithms are proposed for polynomial and differential polynomial systems. The key idea is to eliminate one algebraic variable from a set of polynomials in one step using the theory of multivariate resultant. This method was introduced by Yu Grigor'ev to give a quantifier elimination algorithm with nice computational complexity \cite{grigorev}. In this paper, by adapting this elimination method, we give triangular decomposition algorithms for polynomial and ordinary differential polynomial systems.
In the case of polynomial systems, the algorithm gives an unmixed decomposition and has double exponential complexity.
In the case of differential polynomial systems, the algorithm gives an unmixed radical decomposition and has triple exponential complexity.
This seems to be the first differential triangular decomposition algorithm  with elementary computation complexity.

The rest of this paper is organized as follows. In Section 2, we give a new triangular decomposition algorithm for polynomial systems. In Section 3, we give a new triangular decomposition algorithm for ordinary differential polynomial systems. In Section 4, a summary is given.

\section{Decomposition of algebraic polynomial system}
\label{sec-poly}

In this section,  we give an
algorithm which for given polynomials $h_1, \ldots,  h_k \in\K[x_1, \ldots, $ $ x_n]$,  gives the decomposition $\Zero(h_1, \ldots, h_k)=\cup_q \Zero(\sat(\A_q))$,  where $\A_q$ is a regular triangular set for each $q$. Furthermore,  the computational complexity of the algorithm is given.

\subsection{Basic definition and property}

Let $\K$ be a field of characteristic 0,  and $x_1<x_2< \ldots <x_n$  ordered variables. For every $i\in \{1, \ldots , n\}$,  we define $\K_i=\K[x_1, \ldots , x_i]$ to be the ring of multivariate polynomials in
the variables $x_1, \ldots,  x_i$ with coefficients in $\K $. We write $\deg(f, x_i)$ for the degree of $f$  in $x_i$,
 and $\deg_{x_{i_1}, x_{i_2}, \ldots, x_{i_t}}(f)$ for the degree of $f$ as the multivariate polynomial in $x_{i_1}, x_{i_2}, \ldots , x_{i_t}$.
We call the leading variable of $f$,  denoted by $\lv(f)$,  the greatest
variable $v \in \{x_1, \ldots , x_n\}$ such that $\deg(f, v)> 0 $.

Assuming $\lv(f)=x_i$,  we call $i$ the class of $f$,  denoted by $\cls(f)$.
Regarding $f$ as a univariate polynomial in $\K_{i-1}[x_i]$,  we can write $f=cx_i^d+r$. We call $d=\deg(f, x_i)$
the leading degree of $f$,  denoted by $\ldeg(f)$,  and $c$ the initial of $f$,  denoted by $\ini(f)$ or $I_f$.
%

For Let $\PS$ be a polynomial set and $D$ a polynomial in $\K_n$. For an algebraic closed extension field $\E$ of $\K$,  let
$$\zero(\PS/D)=\{\eta\in \E^n\, |\, \forall P\in \PS,  P(\eta)=0 \wedge D(\eta)\ne0\}.$$

A subset $\T$ of $\K_n$ is called a triangular set if no element of $\T$ lies in $\K$ and for $P, Q\in \T$ with $P\neq Q$ we have $\lv(P)\neq \lv(Q)$.

Let $\T=\{T_1, \ldots, T_r\}$ be a triangular set. We always assume $\lv(T_1) < \lv(T_2) < \cdots < \lv(T_r)$.
We can rename the variables as $u_1, \ldots, u_q, y_1,
\ldots, y_r$ such that $q+r=n$ and $\lv(T_i)=y_i$. Then$\T$ has the following form:
\begin{equation}\T=
\left\{
\begin{array}{lllllllllll}
T_1(u_1, \ldots, u_q, y_1) \\
T_2(u_1, \ldots, u_q, y_1, y_2) \\
\ldots \\
T_r(u_1, \ldots, u_q, y_1, \ldots, y_r) \\
\end{array}
\right\}
\end{equation}
We call $\textbf{u}=\{u_1, \ldots, u_q\}$ the parameter set of $\T$,  and  write $I_{\T}=I_{T_1}\ldots I_{T_r}$.
For a triangular set $\T$,  the saturation ideal of $\T$ is defined to be
 $$\sat(\T)=\{f\in \K_n\, |\, \exists d\in \N^+, \ s.t. \,  I_{\T}^df\in (\T)\}$$
 where $(T)$ is the ideal generated by $\T$.

A triangular set $\T=[T_1, \ldots , T_r]$ of form $(1)$ is called regular,  if for each $1\leq i\leq r$,  $(T_1, \ldots , T_{i-1}, \ini(T_i))\bigcap \K[\textbf{u}]\neq \{0\}$ where $(T_1, \ldots , T_{i-1}, \ini(T_i))$
is the ideal generated by $T_1, \ldots , T_{i-1}, \ini(T_i)$ and
$\textbf{u}$ is the parameter set of $\T$.

\begin{lemma}\cite{dim}\label{prime decomposition}
For a triangular set $\T$, we have:
$$\sqrt{\sat(\T)}=\bigcap_{i=1}^t \sat(\T_i)$$
where $\T_i$ are regular triangular sets having the same parameter set as $\T$, and $\sat(\T_i)$ is a prime ideal.
That is,  $\sat(\T)$ is an unmixed ideal.
\end{lemma}

\begin{lemma}\label{reversible}
Let $\T=\{T_1, \ldots , T_r\}$ be a regular triangular set,  $\textbf{u}$ its parameter set,  $y_i$ the leading variable of $T_i$,  $P$ a polynomial in $\K[\textbf{u}, y_1, \ldots,  y_r]$.
Then $(P, \T)\bigcap \K[\textbf{u}]\neq \{0\}$ if $P$ is not identically zero on all irreducible components of $\zero(\sat(\T))$
\end{lemma}
\proof According to Lemma \ref{prime decomposition},  $\sqrt{\sat(\T)}=\bigcap\limits_{i=1}^t \sat(\G_i)$. Since $P$ is not identically zero on all irreducible component of $\zero(\sat(\T))$,
we have $P\notin \sat(\G_i)$ for each $i$. Since $\sat(\G_i)$ is prime,  so $(P, \G_i)\bigcap \K(\textbf{u})\neq \{0\}$ for each $i$. Suppose that $\G_i=(G_{i, 1}, \ldots , G_{i, r})$ ,
then we have the following equalities:
\begin{eqnarray*}
&&S_{1, 1}G_{1, 1}+\cdots +S_{1, r}G_{1, r}=A_1P+h_1(\textbf{u}) \\
&&S_{2, 1}G_{2, 1}+\cdots +S_{2, r}G_{2, r}=A_2P+h_2(\textbf{u}) \\
&&\quad\cdots \\
&&S_{t, 1}G_{t, 1}+\cdots +S_{t, r}G_{t, r}=A_tP+h_t(\textbf{u}).
\end{eqnarray*}
 Multiply all the equalities. Since the left hand side of the $i$-th equality belongs to $\sat(\G_i)$,  the product of them belongs to $\sqrt{\sat(\T)}$. The product of the right hand side is of the form $AP+h$ for $h=h_1h_2\cdots h_t$. Then
  we have $h\in (P, \sqrt{\sat(\T)})\bigcap \K[\textbf{u}]\neq \{0\}$. Therefore,  there exists an
 integer $d_1$ such that $(h)^{d_1}\in (f, \sat(\T))$. There exists an integer $d_2$ s.t. $(\ini(T_1)\ldots \ini(T_r))^{d_2}(h)^{d_1}\in (P, \T)$.
Since $\T$ is regular,  we have $(\ini(T_i), \T)\bigcap \K[\textbf{u}]\neq \{0\}$, and then the following equality
 $$(\ini(T_1)\ldots \ini(T_r))^{d_2}F_0=g(\textbf{u})+F_1T_1+\ldots +F_rT_r$$
 where $g\ne0$ and $g\in\K[\textbf{u}]$.
Hence,  $(h)^{d_1}g\in (P, \T)$  and $(P, \T)\bigcap \K[\textbf{u}]\neq \{0\}$. \qedd

As a consequence of Lemma \ref{reversible},  we have:

\begin{cor}\label{irreducible condition}
A triangular set $\T=\{T_1, \ldots , T_r\}$ is regular if for each $2\leq i\leq r$,  $\ini(T_i)$ is not identically zero on all irreducible components of $\sat(T_1,\ldots,T_{i-1}).$

\end{cor}

\begin{lemma}\label{zariski closure lemma}
Let $\T=\{T_1, \ldots , T_r\}$ be a regular triangular set with parameter set $\textbf{u}$. If $M\in \K[\textbf{u}]$ ,  then
$\overline{\zero(\T/I_{\T})}=\overline{\zero(\T/MI_{\T})}, $
where $\overline{\mathbb{S}}$ is the Zariski closure of\ $\mathbb{S}$.
\end{lemma}
\proof We first prove the lemma when $\sat(\T)$ is prime. Introduce a new variable $z$ and let $I=I_{\T}$. We have
$$\zero(\T/MI)=\zero(\T, MIz-1)\cap \E^n.$$
So for any $f\in \mathbb{I}(\zero(\T/MI))$ ($\mathbb{I}(V)$ is the ideal of polynomials which vanish on $V$),
$f\in \sqrt{(\T, MIz-1)}$. Let $z=\frac{1}{MI}$,  then there exists an integer $d$ such that $(MIf)^d\in (\T)$,  so we have
$(Mf)^d\in \sat(\T)$. Since $\sat(\T)$ is prime and $M\in \K[\textbf{u}]$ so not in $\sat(\T)$,  we have $f\in \sat(\T)$. So we have
$\overline{\zero(\T/MI)} \supset \zero(\sat(\T))=\overline{\zero(\T/I)}.$
It is obvious $\overline{\zero(T_1, \ldots , T_r/MI)} \subset \overline{\zero(\T/I)}$,  so we have
$\overline{\zero(\T/MI)} = \overline{\zero(\T/I)}.$
Now assuming $\sat(\T)$ is not prime. According to Theorem 1.3 in \cite{dim},  we have
$ \zero(\T/I)=\bigcup\limits_{i=1}^t \zero(\G_i/I_i)$,
where $\G_i$ is a regular triangular set having the same parameter set with $\T$ and $\sat(\G_i)$ is prime for each $i$. Then we have
\begin{eqnarray*}
\overline{\zero(\T/MI)}=&\bigcup\limits_{i=1}^t \overline{\zero(\G_i/I_iM)}
=\bigcup\limits_{i=1}^t \overline{\zero(\G_i/I_i)}
=\overline{\zero(\T/I)}
\end{eqnarray*}
and the lemma is proved. \qedd

\begin{lemma}\label{zariski closure}
Let $\T=\{T_1, \ldots , T_r\}$ be a regular triangular set. If $M$ is not identically zero on all irreducible components of $\sat(\T)$,  then
$\zero(\sat(\T))=\overline{\zero(\T/I_{\T})}=\overline{\zero(\T/MI_{\T})}, $
where $\overline{\mathbb{S}}$ is the Zariski closure of $\mathbb{S}$.
\end{lemma}
\proof
It is well known that $\zero(\sat(\T))=\overline{\zero(\T/I_{\T})}$.
We have $\overline{\zero(\T/I_{\T})}\supset \overline{\zero(\T/MI_{\T})}$,
since $\zero(\T/I_{\T})\supset \zero(\T/MI_{\T})$.
Let $\textbf{u}$ be the parameter set of $\T$,  since $M$
is not identically zero on all irreducible components of $\T$,  according to Lemma \ref{reversible},  we have $(M, \T)\bigcap \K[\textbf{u}]\neq \{0\}$. Suppose
\begin{equation}\label{eq-2}
AM+A_1T_1+\ldots+A_rT_r=H(\textbf{u})
\end{equation}
Let $\xi$ be a zero of $\zero(\T/HI_{\T})$. Substitute $\xi$ into \bref{eq-2},
we have $M(\xi)\neq 0$,
so $\xi$ is also a zero of $\zero(\T/I_{\T})$
and we have
$\zero(\T/MI_{\T})\supset \zero(\T/I_{\T})$.
Since $H\in \K[\textbf{u}]$,  so according to Lemma \ref{zariski closure lemma} we have
$\overline{\zero(\T/HI_{\T})}=\overline{\zero(\T/I_{\T})}$.
Therefore,   we have
$ \overline{\zero(\T/MI_{\T})}\supset \overline{\zero(\T/HI_{\T})}
 =\overline{\zero(\T/I_{\T})}$.
This completes the proof. \qedd

\subsection{A quasi GCD algorithm}

We need to use lemma 1 of \cite{grigorev},  which is modified slightly to the following form.
\begin{lemma}\cite{grigorev}\label{gcd}
There is an algorithm which for given polynomials
$h_i=\sum h_{i, j}Y^j\in \K_n[Y], $ $\deg_{x_1, \ldots, x_n, Y}(h_i)<d,  i=0, 1, \ldots, k$,
yields such two families of polynomials
$g_{q, t}\in \K_n , \Psi _q\in \K_n[Y]$ for $1\leq q\leq N_1 , 0\leq t\leq N_2$ such that
\begin{eqnarray*}
\zero(h_1, \ldots , h_k/ h_0)&=&\bigcup\limits_{q=1}^{N_1} \zero(\Psi_q , g_{q, 1}, \ldots , g_{q, N_2}/ g_{q, 0})\\
&&\bigcup \zero(\{h_{i, j}, i=0, \ldots, k, j=0, \ldots, d-1\}/ h_0).
\end{eqnarray*}
Furthermore, we have the following properties.

$(1).\ \deg(\Psi_q, Y)> 0,  \ini(\Psi_q) \mid g_{q, 0}$.

$(2).\ \deg_{x_1, \ldots , x_n, Y}(\Psi_q) , \deg_{x_1, \ldots , x_n}(g_{q, t})\leq \mathcal P(d);N_1, N_2\leq k\mathcal P(d^n)$ where $\mathcal P(k)$ means a polynomial in $k$.

$(3).\ $ The running time of the algorithm can be bounded by a polynomial in $k$ and $d^n$.
\end{lemma}

Now we describe the main steps of this algorithm without proof. One can refer to \cite{grigorev} for more details.
Without loss of generality,  we assume that $\deg_Y(h_i)>0$ for $0\leq i\leq k$.

Since $\deg_{x_1, \ldots, x_n, Y}(h_i)<d$,  we have $h_i=\sum\limits_{j=0}^{d-1}h_{i, j}Y^j$. Let
$\tilde{h}_{i, j}=\sum\limits_{\beta=0}^j h_{i, \beta}Y^{\beta}$ and
 $$U_{i, j}=\zero(h_{1, d-1}, \ldots, h_{1, 0}, h_{2, d-1}, \ldots, h_{2, 0}, \ldots, h_{i, d-1}, \ldots,
h_{i, j+1}/h_{i, j})$$
 for $1\leq i\leq k$, $0\leq j\leq d-1$. Let $H=\{h_1,\ldots,h_k\}$,
\begin{eqnarray}
H_{i, j}&=&\{\tilde{h}_{i, j}, h_{i+1},\ldots,h_k\}\label{eq-5}\\
H_{k+1}&=&\{h_{i, j},\ \forall\, 1\leq i\leq k\ and\ j\}.\nonumber
\end{eqnarray}
Then we have $\zero(H/h_0)=\zero(H_{k+1}/h_0)\bigcup (\bigcup\limits_{1\leq i\leq k, 0\leq j\leq d-1} \zero(H_{i, j}/h_0)\cap U_{i, j})$.

Now we turn to the system $H_{i, j}$ and introduce new variables $Y_0, Y_1$ to make polynomials in \bref{eq-5} homogeneous in $Y, Y_0, Y_1$. Let
\begin{eqnarray*}
 \overline{h}_i &=& Y_0^j\tilde{h}_{i, j}(x_1, \ldots, x_n, \frac{Y}{Y_0})\\
 \overline{h}_l &=& Y_0^{\ldeg(h_l)}h_l(x_1, \ldots, x_n, \frac{Y}{Y_0}), i+1\leq l\leq k\\
 \overline{h}_0 &=& Y_0^{\ldeg(h_0)+1}(\frac{Y_1}{Y_0}h_0(x_1, \ldots, x_n, \frac{Y}{Y_0})-1)
\end{eqnarray*}
Then $\overline{h}_0, \overline{h}_i, \ldots, \overline{h}_k$ are homogeneous polynomials in $Y, Y_0, Y_1$.
%
%
The solutions of the following homogenous system correspond bijectively to that of \bref{eq-5} except $(1:0:0)$.
\begin{equation}\label{eq-7}
\overline{h}_i=\overline{h}_{i+1}=\ldots=\overline{h}_k=\overline{h}_0.
\end{equation}
Here $\overline{h}_0, \overline{h}_i, \ldots, \overline{h}_k$ are considered as polynomials in $Y, Y_0, Y_1$.

Introduce new variables $U_0, U, U_1$ and let $h_{k+1}=Y_0U_0+YU+Y_1U_1$. We rearrange the polynomials $\overline{h}_0, \overline{h}_i, \ldots, \overline{h}_k$
w.r.t the degree in $Y, Y_0, Y_1$ as $g_0, \ldots, g_{k-i+2}$ and $\gamma_0\geq \gamma_1\geq \ldots \geq \gamma_{k-i+2}$ where $\deg_{Y, Y_0, Y_1}(g_s)=\gamma_s$ for $0\leq s\leq k-i+2$.
Since $\deg_{Y, Y_0, Y_1}(h_{k+1})=1$,  we can assume that $g_{k-i+2}$ is $h_{k+1}$.
Let
 $$D=(\sum\limits_{1\leq l\leq \min\{2, k-i+1\}}(\gamma_l-1))+\gamma_0.$$

We construct the Macaulay matrix $A$ as the representation of the linear map
$$\mathcal{A}:\mathcal{H}_0\oplus\ldots \oplus\mathcal{H}_{k-i+2}\rightarrow \mathcal{H}$$
where $\mathcal{H}_l$ (respectively $\mathcal{H}$) is the linear space of homogenous polynomials in $Y, Y_0, Y_1$ of degree $D-\gamma_l$ (respectively $D$) for $0\leq l\leq k-i+2$,
 and
 $$\mathcal{A}(f_0, \ldots, f_{k-i+2})=f_0g_0+\cdots +f_{k-i+2}g_{k-i+2}$$
The matrix $A$ is of size $C_{D+2}^2 \times \sum\limits_{l=0}^{k-i+2}C_{D-\gamma_l+2}^2$ and
can be represented in a form $A=(A^{(num)}, A^{(for)})$, where the elements of the submatrix $A^{(num)}$ do not contain $U, U_0, U_1$. Actually,  $A^{(num)}$ is the
submatrix of $A$ which corresponds to the basis of $\mathcal{H}_0, \ldots, \mathcal{H}_{k-i+1}$ while $A^{(for)}$ corresponds to the basis of $\mathcal{H}_{k-i+2}$.

About the polynomial system \bref{eq-7},  we have the following lemma:
\begin{lemma}\cite{grigorev}\label{homogeneous}
The rank of the matrix $A$ of the polynomial system \bref{eq-7} is $r=C_{D+2}^2$. Let $\Delta$ be a nonsingular $r\times r$ submatrix of
$A$ containing $\rank (A^{(num)})$ columns in $A^{(num)}$. Then
$$\det(\Delta)=c\prod\limits_{i=1}^{D_1}L_i, where\ L_i=\xi_{i, 0}U_0+\xi_{i}U+\xi_{i, 1}U_1\ and\ c\ is\ a\ constant$$
where $(\xi_{i, 0}:\xi_{i}:\xi_{i, 1})$ is a solution of \bref{eq-7} and the number of occurrences of $\xi_{i, 0}U_0+\xi_{i}U+\xi_{i, 1}U_1$ in the product coincides with the multiplicity of the solution $(\xi_{i, 0}:\xi_{i}:\xi_{i, 1})$ of \bref{eq-7}.
\end{lemma}

To find the $\Delta$ in Lemma \ref{homogeneous}, we use a variant of Gaussian algorithm which will
compute a series of
\begin{equation}\label{ws}\mathcal{W}_s=\{\textbf{x}\in\K^n:P_1=\ldots=P_{s-1}=0, P_s\neq 0\}
\end{equation}
where $ P_1, \ldots, P_s$ are polynomials in $\textbf{x}, U, U_0, U_1$ and linearly independent.
For $\textbf{x}\in \mathcal{W}_s\cap U_{i, j}$,  the determinant \begin{equation}\label{wsd}\Delta_s=\sum\limits_{i=0}^{D_2}E_s^{(i)}U_0^{D_2-i}
\end{equation}
is what we want. For more details about the variant Gaussian algorithm,
one can refer to \cite{grigorev}.

Now we introduce the following quasiprojective varieties
\begin{equation}\label{wsl}\mathcal{W}_s^{(l)}=\{\textbf{x}\in \mathcal{W}_s:E_s^{(0)}=\ldots=E_s^{(l-1)}=0, E_s^{(l)}\neq 0\}
\end{equation}
where $E_s^{(0)}, \ldots, E_s^{(l-1)}$ are polynomials in $\textbf{x}, U, U_1$.
In \cite{grigorev},  it is proved that if we substitute $U_1=0, U=-1, U_0=Y$ into $\frac{\Delta_s}{E_s^{(l)}}$,  and denote the polynomial as $\Psi_s$,
then for each point $\textbf{x}\in \mathcal{W}_s^{(l)}\cap U_{i, j}$,  the solution of $\Psi_s$ as a polynomial in $Y$ is the solution of the polynomial system
\bref{eq-5}. Since the quasiprojective varieties $\mathcal{W}_s^{(l)}\cap U_{i, j}$ can be divided into a series of polynomial systems $V_t=\zero(g_{t, 1}, \ldots, g_{t, N_2}/g_{t, 0})$,
we have
 $$ \zero(H_{i, j})\cap U_{i, j}=\bigcup\limits_t \zero(\Psi_t, g_{t, 1}, \ldots, g_{t, N_2}/g_{t, 0}).$$
If $\Psi_t=1$,  we can delete that component and finally obtain the decomposition in Lemma \ref{gcd}.

Now we write this procedure as an algorithm to be used in the rest of the paper.
\begin{algorithm}[H]\label{alg-gcd-algorithm}
  \caption{\bf --- Quasi GCD Algorithm} \smallskip
\Inp{$\{\{h_1, \ldots,  h_k\}, h_0, Y\}\ where\ h_0, h_1, \ldots,  h_k \in \K_n[Y]$ and $h_i=\sum h_{i, j}Y^j$ for $i=1, \ldots, k, j=0, \ldots, d-1$.}\\
\Outp{$\D=\{\T_0, \ldots , \T_{N_1} \}$,
where $\T_0=\{\{\}, \{h_{i, j}, 1\leq i\leq k, 0\leq j< d\}, \{h_0\}\}$,  $\T_q=\{\{\Psi_q\}, \{g_{q, 1}, \ldots , g_{q, N_2}\}, \{g_{q, 0}\}\}(1\leq q\leq N_1)$,  such that
 $$\zero(h_1, \ldots, h_k/h_0)=\bigcup\limits_{q=0}^{N_1}\zero(\Psi_q, g_{q, 1}, \ldots, g_{q, N_2}/g_{q, 0})$$
where $\Psi_0=0$,  $\deg(\Psi_q, Y)>0$,  and $\deg(g_{q, i}, Y)=0$ for $1\leq q\leq N_1$,  $0\leq i\leq N_2$.}
%
%
\end{algorithm}

\begin{example}
We use a simple example to explain the algorithm. Let the original polynomial system be $\{Y^2+Y/Y\}$. First,  we introduce a new variable $Y_1$ and get an
equivalence system $\{Y^2+Y, Y_1Y-1\}$. Second,  we introduce a new variable $Y_0$ to make it homogeneous $\{Y^2+YY_0, Y_1Y-Y_0^2\}$.
Finally,  we introduce $U, U_0, U_1$ and add $YU+Y_0U_0+Y_1U_1$ to the homogeneous system. The
matrix $A$ corresponding to the homogeneous system is
 \[A =\left [\begin{array}{llllllllllll}
1 &0 &0 &0 &0 &0 &U &0 &0 &0  &0 &0      \\
0&0 &0 &0 &-1 &0 &0 &U_0 &0 &0 &0 &0  \\
0&0 &0 &0 &0 &0 &0 &0 &U_1 &0 &0 &0    \\
0&1 &0 &-1 &0 &0 &0 &U &0 &U_0 &0 &0   \\
0&0 &0 &0 &0 &1 &0 &0 &U &0 &U_1 &0  \\
0& 0&0 &0 &0 &0 &0 &0 &U_0 &0 &0 &U_1  \\
1&1 &0 &0 &0 &0 &U_0 &0 &0 &U &0 &0  \\
0&0 &1 &1 &0 &0 &U_1 &0 &0 &0 &U &0  \\
0&0 &0 &0 &0 &-1 &0 &U_1 &0 &0 &0 &U_0  \\
0&0 &1 &0 &1 &0 &0 &0 &0 &U_1 &U_0 &U  \\
\end{array}\right]\]
$A^{(num)}$ is the submatrix of $A$ formed by the first 6 columns,  $rank(A^{(num)})=6$. According to Lemma \ref{homogeneous},  we must choose the first 6 columns and
by calculating we find the submatrix formed by the first 9 columns and the last column is nonsingular,  which is

\[\Delta =\left [\begin{array}{llllllllllll}
1 &0 &0 &0 &0 &0 &U &0 &0  &0      \\
0&0 &0 &0 &-1 &0 &0 &U_0 &0  &0  \\
0&0 &0 &0 &0 &0 &0 &0 &U_1 &0    \\
0&1 &0 &-1 &0 &0 &0 &U &0  &0   \\
0&0 &0 &0 &0 &1 &0 &0 &U  &0  \\
0& 0&0 &0 &0 &0 &0 &0 &U_0  &U_1  \\
1&1 &0 &0 &0 &0 &U_0 &0 &0  &0  \\
0&0 &1 &1 &0 &0 &U_1 &0 &0  &0  \\
0&0 &0 &0 &0 &-1 &0 &U_1 &0 &U_0  \\
0&0 &1 &0 &1 &0 &0 &0 &0  &U  \\
\end{array}\right]\]
$\det(\Delta)=-U_1^3(U-U_0+U_1)$. Substituting $U_1=0, U=-1, U_0=Y$ to $U-U_0+U_1$,  we obtain the polynomial $-1-Y$ and
$\zero(Y^2+Y/Y)=\zero(Y+1)$

\end{example}


The components of Lemma \ref{gcd} may be empty,  as shown by the following example.
\begin{example}
Let $h_1=xy+1 , h_2=x$,  and take $y$ as the maximal variable. According to Lemma \ref{gcd},  it can be divided into two components $\zero(1, x)$ and $\zero(xy+1, x/x)$.
We can delete the first component. However,  we cannot delete the second component $\zero(xy+1, x/x)$ which is empty. The second component will be deleted in our main algorithm later when we continue our procedure to $\zero(x/x)$.
\end{example}


\subsection{The decomposition algorithm}
We now give the main result about polynomial systems.
\begin{theorem}\label{main theorem}
For a given polynomial system $H=\{h_1, \ldots , h_k\} \in \K_n$,
$\deg(h_i)< d$ for $1\leq i\leq k$,
there is an algorithm to compute regular triangular sets $\mathcal{A}_q=[\Psi_{q, 1}, \ldots , \Psi_{q, l_q} ]$ which have the following properties:

1. $\zero(H)=\cup_{q=1}^N \zero(\sat(\mathcal{A}_q))$ is an unmixed decomposition.

2. The degrees of  $\Psi_{q, 1}, \ldots, \Psi_{q, l_q}$ are less than $d^{c^n}$, $N\leq k^nd^{nc^{n+2}}$,  where $c$ is a constant.

3. The running time of the algorithm can be bounded by a polynomial in $k^n$ and $d^{nc^{n+2}}.$
\end{theorem}

Using the algorithm described below,  we can calculate the regular triangular sets $\A_q$ which satisfy
the properties in Theorem \ref{main theorem}.

\begin{algorithm}[H]\label{alg-satZ1}
  \caption{\bf --- Algebraic Triangular Decomposition} \smallskip
\Inp{$\{h_1, \ldots,  h_k\}, \ where\ h_1, \ldots,  h_k\in \K[x_1, \ldots , x_n]$.}\\
\Outp{$\R$,  which is the set of $(\A_q, D_q)$ and $\A_q=\{\Psi_{q, 1}, \ldots \Psi_{q, l_q}\}$.
$\Psi_{q, i}, D_q\in \K_n$ for $1\leq q\leq N, 1\leq i\leq l_q$ such that
$\A_q$ are regular triangular sets,  $\ini(\Psi_{q, i})\neq 0$ on any element of $\zero(\A_q/D_q)$,  and
 $\zero(h_1, \ldots,  h_k)=\cup_q \zero(\sat(\A_q)).$
}
\medskip
\noindent
1. Let $\T=\{\{\}, \{h_1, \ldots,  h_k\}, \{\}\}$,  $\mathbb{S}=\{\T\}$,  $\R=\{\}$.\\
2. If $\mathbb{S}= \emptyset$ output $\R$,  else let $\T=\{\F, \P, \N\}\in \mathbb{S}$,  and $\mathbb{S}=\mathbb{S}\setminus \{\T\}$.\\
3. If $|\F|>k$,  go to step 2.\\
4. If $\P=\emptyset$,  add $(\F, \prod_{p\in\N} p)$ to $\R$ and go to step 2.\\
5. Let $x_\gamma =\max_{h\in \P}\lv(h)$,
 $\widetilde{\P}=\{h\in \P\, |\, \lv(h)=x_\gamma\}$,  $\P=\P\setminus \widetilde{\P}$.\\
6. Let $\widetilde{\N}=\{f\in \N\, |\, \lv(f)\leq x_\gamma\}$,  $H=\prod\limits_{f\in \widetilde{\N}}f$.\\
7. Apply Algorithm 1 to $\{\widetilde{\P}, H, x_\gamma\}$,  and let the output be $\D$.\\
8. If $\D= \emptyset $,  go to step 2,  else let $\T_1=\{\mathbb{W}, \mathbb{U}, \mathbb{V}=\{v\}\}\in \D$, $\D=\D\setminus \{\T_1\}$.\\
9.
 Let
 $\mathbb{U}=\mathbb{U}\cup \P$,  $x_\eta =\max_{f\in \mathbb{U}}\lv(f)$.\\
10. If $\lv(v)\leq x_\eta$ or $\U=\emptyset$,  add $\{\F\cup\mathbb{W},  \mathbb{U}, \N\cup\mathbb{V}\}$ to $\mathbb{S}$.\\
11. If $\lv(v)> x_\eta$,  write $v=\Sigma l_{\alpha}\textbf{x}^{\alpha}$ as a multivariate polynomial in $\textbf{x}=(x_\eta, \ldots , x_\gamma)$
 with coefficients in $\K[x_1, \ldots, x_{\eta-1}]$. Add
$\{\F\cup \mathbb{W}, \mathbb{U}, \N\cup\mathbb{V} \cup \{l_\alpha\}\}$ to $\mathbb{S}$ for each $\alpha$. Go to step 8.
\end{algorithm}
\begin{example}
A simple example is used to explain the algorithm. Let $f=xyz+1, g=x^2+x$,  $x<y<z$.
In step 5,  $x_\gamma=z$ and $\tilde{\mathbb{P}}=\{f\}$. In step 7, applying Algorithm 1  to $\tilde{\mathbb{P}}$,
the output is $\D_1=\{\T_1\}$ where $\T_1=\{\{xyz+1\}, \{\}, \{xy\}\}$. In step 9,  we have $\U=\{x^2+x\}$,  $x_\eta=x$. Since $lv(xy)=y>x$,  we execute step 11 and add $\{\{xyz+1\}, \{x^2+x\}, \{xy, x\}\}$ to $\mathbb{S}$
and go to step 2. Now we have $\T=\{\F, \P, \N\}$,  where $\F=\{xyz+1\}$,  $\P=\{x^2+x\}$, $\N=\{xy, x\}$.
In step 5,  we have $x_\gamma=x$,  $\tilde{\P}=\{x^2+x\}$. In step 6,  we have $\tilde{\N}=\{x\}$,  $H=x$.
 Applying Algorithm 1 to $\{\tilde{\P}, H,x\}$,  the output is $\{\{x+1\}, \{\}, \{\}\}$. In step 10,  we add $\{\{xyz+1, x+1\}, \{\}, \{xy\}\}$ to $\mathbb{S}$. In step 4,  since $\P=\emptyset$,  we add
$\{\{xyz+1, x+1\}, \{xy\}\}$ to $\R$ and output $\R$. Finally we have
$$\zero(xyz+1, x^2+x)=\zero(xyz+1, x+1/xy)=\zero(\sat(xyz+1, x+1)).$$
The purpose of step 11 is to add $x$ to $\N$. Otherwise, we will apply Algorithm 1 to $\{\{x^2+x\}, xy,x\}$, which does not satisfy the input condition of Algorithm 1 since $x < y$.
\end{example}

Before proving Theorem \ref{main theorem},  we first prove several lemmas.

\begin{lemma}\label{lm-1}
Algorithm 2 terminates and each $\A_q$ is a triangular set.
\end{lemma}
\proof By Lemma \ref{gcd},  after step 7,  for any $\{\mathbb{W}, \mathbb{U}, \mathbb{V}\}\in \D$,  we have $\lv(p) < x_\gamma$ for any $p\in \mathbb{U}$. In other words, for any new $\{\F, \P, \N\}$ added to $\mathbb{S}$ in steps 10 and 11,  the class of the polynomials in $\P$ will be decreased at least  by one. Therefore,  the algorithm terminates. Also,  $\mathbb{W}$ is either empty or  $\mathbb{W}=\{p\}$ and $\lv(p)=x_\gamma$,  which means $\A_q$ is a triangular set for each $q$.\qedd

\begin{lemma}\label{lm-2}
Omitting Step 3, $\zero(h_1, \ldots,  h_k)=\cup_{q=1}^N \zero(\A_q/D_q)$,
and $\ini(\Psi_{q, i})\neq 0$ on any element of $\zero(\A_q/D_q)$.
\end{lemma}
\proof
To show $\zero(h_1, \ldots,  h_k)=\bigcup\limits_q \zero(\A_q/D_q)$,  it suffices to show that the equality
\begin{equation}\label{eq-po1}
\zero(h_1, \ldots,  h_k)=\cup_{\{\F, \P, \N\}\in\mathbb{S}}\zero(\F\cup\P/\prod_{p\in\N} p)
\end{equation}
always holds in the algorithm,  and when $\P=\emptyset$ the algorithm returns the requires equation.
$\mathbb{S}$ is modified in steps 7, 10 and 11.
In step 7,  by Lemma \ref{gcd},  $\zero(\widetilde{\P}/H) = \cup_{\{\mathbb{W}, \mathbb{U}, \{v\}\}\in \D}\zero(\{\mathbb{W}\cup\mathbb{U}/v)$. Clearly,  after applying  Algorithm 1,  \bref{eq-po1} remains valid when $\widetilde{\P}$ and $\widetilde{\N}$ are properly replaced as in steps 10 and 11.
In step 1,  a special substitution is performed.
Let $v=\Sigma l_{\alpha}\textbf{x}^{\alpha}$. Then $\zero(/v)$ is replaced by $\cup_{\alpha}\zero(/l_{\alpha}v)$. Since $\zero(/v)=\cup_{\alpha}\zero(/l_{\alpha}v)$,  \bref{eq-po1} is still valid after step 11.

Now suppose $(\Psi_{1}, \ldots , \Psi_{t}/M)$ is one component of the output. From the procedure of the algorithm,  we know
that this component is obtained in the following manner:
\begin{eqnarray}
\zero(f_{0, 1}, \ldots , f_{0, k^{(0)}}/M_0)
 &\rightarrow&\zero(\Psi_1, f_{1, 1}, \ldots , f_{1, k^{(1)}}/M_1)\nonumber\\
 &\rightarrow&\zero(\Psi_1, \Psi_2, f_{2, 1}, \ldots , f_{2, k^{(2)}}/M_1M_2)\label{eq-pr1}\\
 &\rightarrow&\ldots\nonumber\\
 &\rightarrow&\zero(\Psi_1, \Psi_2, \ldots , \Psi_t/M_1\ldots M_t)\nonumber
\end{eqnarray}
and $M=M_1\ldots M_t$.
Note that after applying Algorithm 1,
$\zero(\Psi_1, f_{1, 1}, \ldots , f_{1, k^{(1)}}/T_1)$ is a component of $\zero(f_{0, 1}, \ldots , f_{0, k^{(0)}})/M_0)$. If $\lv(v)\le x_\eta$
in step 10,  $M_1=T_1$. Otherwise,
$M_1$ is the multiplication of $T_1$ and a coefficient $l_{\alpha}$ of $T_1$ as shown in step 11.
The component $\zero(\Psi_2, f_{2, 1}, \ldots , f_{2, k^{(2)}}/M_2)$ is obtained similarly from $\zero(f_{2, 1}, \ldots , f_{2, k^{(2)}}/S_2)$, where $S_2$ is the maximal factor of $M_1$ satisfying $\lv(S_2) \le \lv(f_{2, j})$ for all $j$. Continuing this procedure, we will obtain \bref{eq-pr1}. It is obvious that
 \begin{eqnarray}
\zero(f_{0, 1}, \ldots , f_{0, k^{(0)}}/M_0)&\supset&\zero(\Psi_1, f_{1, 1}, \ldots , f_{1, k^{(1)}}/M_1)\nonumber\\
&\supset&\zero(\Psi_1, \Psi_2, f_{2, 1}, \ldots , f_{2, k^{(2)}}/M_1M_2)\label{eq-pr2}\\
&\supset&\ldots \nonumber\\
&\supset&\zero(\Psi_1, \Psi_2, \ldots , \Psi_t/M_1\ldots M_t)\nonumber
\end{eqnarray}
According to (1) of Lemma \ref{gcd},  we have  $\ini(\Psi_i)\mid M_i$,  so $\ini(\Psi_i)\neq 0$ on any element of $\zero(\Psi_{1}, \ldots ,
 \Psi_{t}/M)$. \qedd

\begin{lemma}\label{lm-3}The triangular sets $\A_q=\{\Psi_{q, 1}, \ldots \Psi_{q, l_q}\}$ are regular and $\overline{\zero(\A_q/D_q)}$
$=\zero(\sat(\A_q))$.
\end{lemma}
\proof
Let $\zero(\Psi_{1}, \ldots , \Psi_{t}/M)$ be a component of the output.
According to the proof of Lemma \ref{lm-2},  this component comes from procedure \bref{eq-pr1}.
Now we assume that
 $\lv(\Psi_1)=x_{k_1},   M_1 \in \K_{k_1-1}$,
 $\lv(\Psi_2)=x_{k_2},   M_2 \in \K_{k_2-1}$, $\ldots$,
 $\lv(\Psi_t)=x_{k_{t}}, M_t\in \K_{k_{t}-1}$.

According to Lemma \ref{irreducible condition}, to show that $\A_q$ is regular,  it suffices to prove that $\ini(\Psi_i)$ is not always zero on any irreducible component of $\sat(\Psi_{i+1}, \ldots , \Psi_t)$ for $1\leq i\leq t-1$.
We prove this by induction.
First,  supposing $\ini(\Psi_{t-1})$ is zero on an irreducible component of $\sat(\Psi_t)$. $\zero(\Psi_t/M_t)$ is a component of $\zero(f_{t-1, 1}, \ldots , f_{t-1, k^{(t-1)}}/S_{t-1})$ after applying Algorithm
1, where $S_{t-1}$ is a factor of $M_{t-1}$. Obviously,  $\zero(\Psi_t/M_t)$ is not empty and $\overline{\zero(\Psi_t/M_t)}=\overline{\zero(\Psi_t/\ini(\Psi_t))}=\zero(\sat(\Psi_t))$ since $\lv(M_t)<\lv(\Psi_t)$.
%
%
Since $\ini(\Psi_{t-1})$ is always zero on an irreducible component of $\sat(\Psi_t)$,  there exists an $\eta_{k_{t}}=(\xi_1, \ldots, \xi_{k_{t}})$ in $\zero(\Psi_t/M_t)$ such that $\ini(\Psi_{t-1})(\eta_{k_{t}})=0$.
Since $\zero(f_{t-1, 1}, \ldots , f_{t-1, k^{(t-1)}}/S_{t-1})
\supset \zero(\Psi_t/M_t)$,  $\eta_{k_{t}}\in\zero(f_{t-1, 1}, \ldots , f_{t-1, k^{(t-1)}}/S_{t-1})$.
If $\zero(\Psi_t/M_t)$ is obtained from step 10,  then  $\eta=\eta_{k_{t}}$ is also in $\zero(f_{t-1, 1},$ $ \ldots , f_{t-1, k^{(t-1)}}/M_{t-1})$.
Otherwise,  $\zero(\Psi_t/M_t)$ is obtained from step 11,  $\eta_{k_{t}}$ can be extended to a zero $\eta=\eta_{k_{t-1}-1}$ of $\zero(f_{t-1, 1}, \ldots , f_{t-1, k^{(t-1)}}/M_{t-1})$, since $S_{t-1}$ is a coefficient of $M_{t-1}$. So in each case $M_{(t-1)}(\eta)$$\neq 0$, but we have
$\ini(\Psi_{t-1})|M_{t-1}$, a contradiction. We have proved $\{\Psi_{t-1}, \Psi_{t}\}$ is regular.
We can prove in the same way that $M_{t-1}$ is not always zero on any irreducible component of $\sat(\Psi_t)$. According to Lemma \ref{zariski closure},  we have
$\overline{\zero(\Psi_{t-1}, \Psi_{t}/\ini(\Psi_t)\ini(\Psi_{t-1}))}
=\overline{\zero(\Psi_t, \Psi_{t-1}/M_{t-1}M_t)}.$
The induction step can be proved similarly.\qedd

\begin{lemma}\label{lm-4}
In Algorithm 2,  the degree of the polynomials  $\Psi_{q, 1}, \ldots , \Psi_{q, l_q}$ are less than $d^{c^n}$ and $N\leq k^nd^{nc^{n+2}} $,  where $c$ is a constant.
The running time of the algorithm can be bounded by a polynomial in $k^n$ and $d^{nc^{n+2}}.$
\end{lemma}
\proof
According to Lemma \ref{gcd},  for given polynomials $h_1, \ldots , h_k\in \K_n$ with  $\deg(h_i)<d$,  after applying Algorithm 1,  we obtain no more than $kd^{cn}$ components,
each component has no more than $kd^{cn}$ polynomials,  the degrees of polynomials in these components are less than $d^c$,  and the running time of the algorithm can be bounded by a polynomial in $k, d^n$.
After applying Algorithm 1,  the most complicated situation is that the maximal leading variable of the polynomials $g_{q, t}$ is $x_{n-1}$. Applying Algorithm 1 to these components,  each component will be split
to at most $kd^{cn}d^{c^2(n-1)}\leq kd^{c^3n}$ components,  each component has at most $kd^{cn}d^{c^2(n-1)}\leq kd^{c^3n}$ polynomials,  and the degree of each polynomial is less than $d^{c^2}$.
This procedure will terminate in at most $n$ steps. In step $n$,  each component will be split to at most $kd^{c^{n+1}n}$ components,  each component has at most $kd^{c^{n+1}n}$ polynomials, and each polynomial has degree less than $d^{c^n}$.
Then in total,  there are at most $k^nd^{c^{n+2}n}$ components,  and the degree of the polynomials can be bounded by $d^{c^n}$. The running time
of Algorithm 2 can be bounded by a polynomial in $k^n, d^{c^{n+2}n} $.  \qedd

{\em Proof of Theorem \ref{main theorem}.}
Omitting Step 3, the correctness of the theorem follows from Lemmas \ref{lm-1}, \ref{lm-2}, \ref{lm-3}, and \ref{lm-4}.
It suffices to show that with step 3, the theorem is also correct. Suppose $\widetilde{\A}_k,k=1,\ldots,N_0$ are the extra regular triangular sets obtained by omitting Step 3 and $\A_l,l=1,\ldots,N$ are those obtained with Step 3.
Then
$$\zero(h_1,\ldots,h_k)=\cup_{l=1}^{N} \zero(\sat(\A_l))\bigcup \cup_{k=1}^{N_0} \zero(\sat(\widetilde{\A}_k)).$$
From the condition $|\F|>k$ in Step 3, we have $|\widetilde{\A}_k|>k$.
By the dimension theorem proved in \cite{dim},  $\dim(\zero(\sat(\widetilde{\A}_k))) < n-k$. While by the affine dimension theorem \cite[p. 48]{hat}, any component of $\zero(h_1,\ldots,h_k)$ is of dimension no less than $n-k$. Thus, $\zero(\sat(\widetilde{\A}_k))$ are redundant in the decomposition and can be deleted.\qedd

\section{Decomposition of ordinary differential polynomial systems}
In this section, a decomposition algorithm for ordinary differential polynomial systems will be given, which has an elementary worst case complexity bound.

\subsection{Basic definition and property}

Let $\K$ be a field of characteristic zero in which an operation of differentiation is performable such that for any $a, b\in \K$,
 $$(a+b)'=a'+b',  (ab)'=ab'+ba'.$$
Then we call $\K$ a differential field. Let $y_1, \ldots, y_n$ be differential indeterminates. We write the $j$-th derivative of $y_i$ as $y_{i}^{(j)}$.
Let $\K\{y_1, \ldots , y_n \}=\K[y_i^{(j)}, i=1, \ldots, n;j\in \N]$ be the ring of differential polynomials in $y_1, \ldots, y_n$.

Let $f$ be a differential polynomial in $\K\{y_1, \ldots , y_n \}$. The class of $f$ denoted by $\cls(f)$,  is the greatest $p$
such that some $y_{p}^{(j)}$ is present in $f$. If $f\in\K$, then $\cls(f)=0$.
The order of $f$ w.r.t $y_i$, denoted by $\ord(f, y_i)$, is the greatest $j$ such that $y_{i}^{(j)}$ appears effectively in $f$.
We write $\ord(f)=\max_{1\leq i\leq n}\ord(f, y_i)$. If $\cls(f)=i$ and $\ord(f, y_i)=j$  then we call $y_{i}^{(j)}$ the leader of $f$,
and we write it as $\ld(f)=y_{i}^{(j)}$. We define $y_i^{(j)}=\ld(f)>\ld(g)=y_\alpha^{(\beta)}$ if $i>\alpha$ or $i=\alpha, j>\beta$.
We can write $f$ as a univariant polynomial in its leader such that $f=a_d(y_{i}^{(j)})^d+\ldots +a_0$,  and we call $a_d$ the initial of $f$,  which is denoted by $I_f$. We call $\frac{\partial f}{\partial y_i^{(j)}}$  the separant of $f$,  which is denoted by $S_f$.
For $f, g \in \K\{y_1, \ldots , y_n\}$,  we say $f$ is of higher rank than $g$,  if one of the following conditions is satisfied

   1. $\cls(f)>\cls(g)$.

   2. $\cls(f)=\cls(g)=p$ and $\ord(f, y_p)>\ord(g, y_p)$.

   3. $\cls(f)=\cls(g)=p,  \ord(f, y_p)=\ord(g, y_p)=j$, and $\deg(f, y_p^{(j)})>\deg(g, y_p^{(j)})$.

Let $\cls(g)=p>0$. We say $f$ is reduced w.r.t $g$,  if $\ord(f, y_p)<\ord(g, y_p)$ or $\ord(f, y_p)=\ord(g, y_p)=j$, and $\deg(f, y_p^{(j)})<\deg(g, y_p^{(j)})$.

For $f_1, \ldots, f_k \in\K\{y_1, \ldots, y_n\}$,  we use $[f_1, \ldots, f_k]$ to denote the differential ideal generated by $f_1, \ldots,  f_k$,  which is  the linear combination of $f_1, \ldots,  f_k$ and their derivatives.

A set $\T:=\{T_1, \ldots,  T_r\}$ of differential polynomials in $\K\{y_1, \ldots , y_n\}$ is called a triangular set,  if $\cls(T_i)\neq \cls(T_j)$ for  $i\neq j$.
Assuming that $\cls(T_1)<\cdots <\cls(T_r)$,  we rename the variables as
$u_1, \ldots , u_t,y_1,\ldots,y_r$ such that $r+t=n$
and $\ld(T_i)=y_i^{(\gamma_i)}$.
A differential polynomial $f\in \K\{u_1, \ldots , u_t,y_1,\ldots,y_r\}$ is said to be invertible w.r.t $\T$ if $[f, T_1, \ldots, T_r]\cap \K\{u_1, \ldots , u_t\}\neq \{0\}.$
$\T$ is called regular if $I_{T_i}$ are invertible w.r.t to $\T_{i-1}$  for $0\leq i\leq r$. $\T$ is called  saturated if $\T$ is regular and $S_{T_i}$ are invertible w.r.t to $T_i$ for $1\leq i\leq r$.

Let $\T:=\{T_1, \ldots,  T_r\}$ be a triangular set.
Denote $I_\T=I_{T_1}\cdots I_{T_r}$ and $S_\T=S_{T_1}\cdots S_{T_r}$.
Then the saturation ideal of $\T$ is
 $$\dsat(T)=\{f\in\K\{y_1, \ldots , y_n\}\,|\,\exists d\in N,  s.t.\,
 (I_\T S_\T)^df\in [T_1, \ldots , T_r]\}.$$
It is known that if $\T$ is saturated, then $\dsat(\T)$ is an unmixed radical differential ideal \cite{boul,rody1}.

\begin{lemma}\label{differential regular}
Let $\T:=\{T_1, \ldots,  T_r\}$ be a triangular set in $\K\{y_1, \ldots , y_n\}$. Then $\T$ is saturated if $I_{T_i}$ and $S_{T_i}$ are not identically zero on all irreducible components of $\dsat(\T_{i-1})$ and $\dsat(\T_{i})$,  for $1\leq i\leq r$,  respectively.
\end{lemma}
\proof This lemma can be proved similar to Lemma \ref{irreducible condition}.\qedd

\begin{lemma}\label{differential zariski}
Let $\T:=\{T_1, \ldots,  T_r\}$ be a saturated triangular set in $\K\{y_1, \ldots , y_n\}$. If $M\in \K\{y_1, \ldots , y_n\}$ is not identically zero
on all irreducible components of $\dsat(\T)$,  then we have
$\overline{\zero(\T/I_\T S_\T)}=\overline{\zero(\T/MI_\T S_\T)}=\zero(\dsat(\T))$.
\end{lemma}
\proof  This lemma can be proved similar to Lemma \ref{zariski  closure lemma}. \qedd


\subsection{A squarefree quai GCD algorithm}

In order to decompose differential polynomial systems, we need to modify Lemma \ref{gcd}. In Lemma \ref{gcd},  for given polynomials $h_0, h_1, \ldots, h_k\in \K[x_1, \ldots, x_n, Y]$, $\deg(h_i)<d$,  we can write
$h_i (i>0)$ as $h_i=\sum\limits_{j=0}^{d-1} h_{i, j}Y^j, $ and divide the whole space as $\K^n=\bigcup\limits_{i, j}U_{i, j}\bigcup\{\textbf{x}\in \K^n\,|\,h_{i, j}(\textbf{x})=0, \forall\ 1\leq i\leq k\ \hbox{ and }\ 0\le j\le d-1\}$,  where
 $$U_{i, j}=\zero(h_{1, d-1}, \ldots, h_{1, 0}, h_{2, d-1}, \ldots, h_{2, 0}, \ldots, h_{i, d-1}, \ldots,h_{i, j+1}/h_{i, j})$$
for $1\leq i\leq k$, $0\leq j\leq d-1$. We write
$\tilde{h}_{i, j}=\sum\limits_{0\leq \beta \leq j}h_{i, \beta}Y^\beta$. Then on $U_{i, j}$, the original polynomial system becomes
\begin{equation}\label{eq-A1}
\tilde{h}_{i, j}=h_{i+1}=\ldots=h_k=0;h_0\neq 0.\end{equation}
We add a step here to divide \bref{eq-A1} into the following polynomial systems:
 \begin{eqnarray}
 &&\tilde{h}_{i, j}=h_{i+1}=\ldots=h_k=0, h_0
 \frac{\partial \tilde{h}_{i, j}}{\partial Y}\neq 0\nonumber\\
 &&\tilde{h}_{i, j}=h_{i+1}=\ldots=h_k=\frac{\partial \tilde{h}_{i, j}}{\partial Y}=0, h_0\frac{\partial^2 \tilde{h}_{i, j}}{\partial Y^2}\neq 0\label{eq-A2}\\
 &&\ldots\nonumber\\
 &&\tilde{h}_{i, j}=h_{i+1}=\ldots=h_k=\frac{\partial \tilde{h}_{i, j}}{\partial Y}=\ldots=\frac{\partial^{j-1} \tilde{h}_{i, j}}{\partial Y^{j-1}}=0, h_0\neq 0.\nonumber
\end{eqnarray}
Since $\frac{\partial^j\tilde{h}_{i, j}}{\partial Y^j}=h_{i, j}$,  and $h_{i, j}\neq 0$ on $U_{i, j}$,
we actually have $\frac{\partial^j\tilde{h}_{i, j}}{\partial Y^j}\neq 0$.
Then the zero set of \bref{eq-A1} equals to the union of the zero sets of \bref{eq-A2}. Now we continues to introduce new variables as in Lemma \ref{gcd} to make the polynomial systems homogenous. After this modification,  Lemma \ref{gcd} becomes the following form.
\begin{lemma}\label{differential gcd}
Given polynomials $h_0, h_1, \ldots,  h_k\in\K_n[Y]$, $\deg(h_i)<d$, and $h_i=\sum_{j=0}^{d-1} h_{i,j}Y^j$ for $0\leq i\leq k$,  we may compute  $g_{q, t}\in \K_n$,  $\Psi_q\in \K_n[Y]\setminus\K_n$ for $1\leq q\leq N_1, 0\leq t\leq N_2$ such that:
$$\zero(h_1, \ldots , h_k/ h_0)=\bigcup\limits_{q=1}^{N_1} \zero(\Psi_q , g_{q, 1}, \ldots , g_{q, N_2}/ g_{q, 0}) \cup
 \zero(\{h_{i, j}, 1\leq i\leq k,0\le j< d \}/ h_0)$$
which has the following properties:

$1.$ We have $\ini(\Psi_q) \mid g_{q, 0}$, and
$S_{\psi_q}\neq 0$ on any element of $\zero(\Psi_q , g_{q, 1},$ $ \ldots,$ $g_{q, N_2}/g_{q, 0})$.

$2.\ \deg_{X_1, \ldots , X_n, Y}(\Psi_q)$, $\deg_{X_1, \ldots , X_n}(g_{q, t})\leq \mathcal P(d);N_1, N_2\leq k\mathcal P(d^n)$.

$3.$ The running time of the algorithm can be bounded by a polynomial in $k , d^n$.
\end{lemma}

\proof For property 1, we need only to prove that  $S_{\psi_q}\neq 0$ on any element of $\zero(\Psi_q , g_{q, 1}, \ldots,$ $g_{q, N_2})$. Since we divide \bref{eq-A1} into the union of \bref{eq-A2},  each component of the output,  for example $(\Psi_1, g_1, \ldots , g_{N_2}/g_0)$,  comes from one of \bref{eq-A2}. Without loss of generality,  suppose it is the first one in \bref{eq-A2}.
Then we have
 $$\zero(\tilde{h}_{i, j},h_{i+1},\ldots,h_k/h_0\frac{\partial \tilde{h}_{i, j}}{\partial Y})\supset \zero(\Psi_1, g_1, \ldots , g_{N_2}/g_0).$$
If $S_{\Psi_1}$ vanishes on $(\xi_1, \ldots, \xi_n, \eta)\in\zero(\Psi_1, g_1, \ldots , g_{N_2}/g_0)$, then $\eta$ must be
a multiple root of $\Psi_1$ when substituting $(x_1, \ldots, x_n)$
by $(\xi_1, \ldots, \xi_n)$. According to Lemma \ref{homogeneous},  $\eta$ is also a multiple root of the homogeneous equation system of
$\bref{eq-A2}$ after introduce new variables $Y_1, Y_0$,  which means $\frac{\partial \tilde{h}_{i, j}}{\partial Y}(\xi_1, \ldots, \xi_n, \eta)=0$,
a contradiction. Property 1 has been proved.

Property 2 comes from Lemma \ref{gcd}. We now prove property 3.
According to the procedure of this algorithm,  the origin system has been divided into no more than $kd$ subsystems $H_{i, j}$ in \bref{eq-5}. For each $H_{i, j}$,  we divide it into no more than $d$
subsystems in \bref{eq-A2}, and each system has no more than $k+d$ polynomials, and the degree of these polynomials are bounded by $2d$. The related matrix $A$ has $C_{D+2}^2$ rows,  where
 $$D=(\sum\limits_{1\leq l\leq \min\{2, k-i+1\}}(\gamma_l-1))+\gamma_0\leq 6d.$$
The degree of the elements in $A$  are bounded by $2d$,  so the degree of $P_s$ in \bref{ws} and $\Delta_s$ in \bref{wsd} are bounded by
$2dC_{D+2}^2\leq d(6d+1)(6d+2)$. Since $P_s$ are linearly independent,  we have $s\leq (d(6d+1)(6d+2))^n)$.
For each  $\mathcal{W}_s$ in \bref{ws},  we divided it into $\mathcal{W}_s^{(l)}$  in  \bref{wsl} and $l$ is no more than the degree of $\Delta$. So in total, we have $kd^2(d(6d+1)(6d+2))^{(n+1)})$ components and
$N_1\leq k\mathcal P(d^n)$.
According to the above proof, it is obvious that the degree of each polynomial in these components is no more than the degree of $P_s$ and $\Delta$,  so is bounded by $\mathcal P(d)$. The polynomials $g_{q, t}$ come from three parts. The first part is the polynomials in $U_{i, j}$ and whose number is bounded by $kd$; the second part is the coefficient of $P_s$ when taken as polynomials in $U, U_0, U_1$ and so the number is bounded by $(d(6d+1)(6d+2))^{2n})$; the third part is the coefficients of $E_s^{(i)}$ and so the number is bounded by $(d(6d+1)(6d+2))^{n+1})$. Therefore, $N_2\leq k\mathcal P(d^n)$. \qedd

Now we write this theorem as an algorithm. We only give the input and output of this algorithm, since the procedure
of this algorithm has been described above.
\begin{algorithm}[H]\label{alg-gcd-d}
  \caption{\bf --- Squarefree Quasi GCD} \smallskip
\Inp{$\{\{h_1, \ldots,  h_k\},\{h_0\}, \{x_1, \ldots , x_n\}, Y\}$ where $h_0, h_1, \ldots,  h_k \in \K[x_1, \ldots , x_n,Y]$, $\deg(h_i)<d$.}\\
\Outp{$\D=\{\T_0, \ldots , \T_{N_1} \}$,
where $\T_0=\{\{\}, \{h_{i, j}, 1\leq i\leq k, 0\leq j\leq d-1\}, \{h_0\}\}$,  $\T_q=\{\{\Psi_q\}, \{g_{q, 1}, \ldots , g_{q, N_2}\}, \{g_{q, 0}\}\}(1\leq q\leq N_1)$, which satisfy the conditions in Lemma \ref{differential gcd}.}
%
%
%
\end{algorithm}

\subsection{The algorithm}
We now give the main result for differential polynomial systems.
\begin{theorem}\label{th-dm}
Let $h_1, \ldots, h_k\in \K\{y_1, \ldots, y_n\}$, where
$\deg(h_i)<d$ and $\ord(h_i)<R$ for $1\leq i\leq k$. There is an algorithm to compute saturated triangular sets $\A_q:=\Psi_{q, 1}, \ldots, \Psi_{q, l_q}$ which have the following properties:

1. $\zero(h_1, \ldots, h_k)=\cup_{q=1}^{N} \zero(\sat(\A_q))$.

2. We have $\deg(\Psi_{q, i})\le d^{c^{2^nR}}$, $\ord(\Psi_{q, i})\le2^nR$,  and $N<k^{2^nR}d^{c^{2^nR}Rn}$.

3. The running time of this algorithm can be bounded by a polynomial in $k^{2^nR}d^{c^{2^nR}Rn}$.
\end{theorem}

We will give an algorithm to produce those saturated triangular sets in the theorem.
Before giving the main algorithm, two sub-algorithms will be given. The first one is the partial remainder \cite{boul,grigorev}.
\begin{algorithm}[H]\label{dprem}
  \caption{\bf --- DPM Algorithm} \smallskip
\Inp{$\{\{g_0\},\{f_1,\ldots,f_k\},\{f_0\}\}$, where $g_0,f_0,\ldots, f_k \in \K\{y_1,\ldots ,y_n\}$, $\ord(f_i,y_\alpha)\leq r$ for $0\leq i\leq k$, and $\ld(g_0)= y_\alpha^{(r-t)},t\geq 1$.}\\
\Outp{$\{\{g_0,\tilde{f}_1,\ldots,\tilde{f}_k\},\{\tilde{f}_0S_{g_0}\} \}$ where $\ord(\tilde{f}_i,y_\alpha)\leq r-t$ for $0\leq i\leq k$ such that
$\zero(g_0,f_1,\ldots,f_k/f_0S_{g_0})=\zero(g_0,\tilde{f}_1,\ldots,\tilde{f}_k/\tilde{f}_0S_{g_0}).$}

1. For $i=0,\ldots,k$,\\
\SPC 1.1.
 $\tilde{f}_i=f_i$.\\
\SPC 1.2.
 If $\ord(\tilde{f}_i,y_\alpha)\le r-t$, goto step 1.\\
\SPC 1.3.
  Let $\ord(\tilde{f}_i,y_\alpha)=r_i$ and $g_0^{(r_i-r+t)}=S_{g_0}y_\alpha^{(r_i)}-H_{r_i}$.\\
\SPC 1.4.
  Replace $y_\alpha^{(r_i)}$ in $\tilde{f}_i$ by $\frac{H_{r_i}}{S_{g_0}}$
 and multiply by $(S_{g_0})^{\deg(\tilde{f}_i,y_\alpha^{(r_i)})}$, and let $\tilde{f}_i$ be the new differential polynomial. Goto step 1.2.\\
2. Output $\{\{g_0,\tilde{f}_1,\ldots,\tilde{f}_k\},\{\tilde{f}_0S_{g_0}\} \}$.
\end{algorithm}

\begin{lemma}\label{prem}\cite{grigorev}
Use the notations in Algorithm 4 and assume $\deg(f_j)<d, \deg(g_0)<d$,
$\ord(f_i, y_\gamma)<R$ for $0\leq i\leq k, 1\leq \gamma\leq n$.
Then,  we have the following bounds: $\ord(\tilde{f}_j, y_\gamma)\leq R+t, \deg(\tilde{f}_j)\leq \mathcal{P}(d, t)$ for any $0\leq j\leq k, 1\leq \gamma\leq n$.
\end{lemma}

Next, we describe a splitting subroutine from \cite{grigorev}.
Let $g\in \K\{y_1, \ldots, y_n\}$. For $\alpha \in\{1, \ldots, n\}$, let $\ord(g, y_\alpha)=r$,
 $g=\sum_a g_{a}(y_\alpha y_\alpha^{(1)}\ldots y_\alpha^{(r)})^a, a=(a_0, a_1, \ldots , a_{r}),
 (y_\alpha y_\alpha^{(1)}\ldots $ $y_\alpha^{(r)})^a$ $=y_\alpha^{a_0}\ldots (y_\alpha^{(r)})^{a_{r}}$. Denote $\coeff(g,y_\alpha)$ to be set of $g_{i, \alpha}$.
For $G\subset \K\{y_1, \ldots, y_n\}$, denote
 $$\coeff(G,y_\alpha) = \cup_{g\in G}\, \coeff(g,y_\alpha).$$
%
%
We have the following split algorithm.
\begin{algorithm}[H]\label{splitting subroutine}
  \caption{\bf --- SPLIT Algorithm} \smallskip
\Inp{$\{G, y_\alpha\}$,  where
$G=\{g_1, \ldots, g_l\}\subset\K\{y_1, \ldots , y_n\}$.}\\
\Outp{$\D= \{\T_0, \ldots, \T_{N}\}$ where
 $\T_0=(\coeff(G,y_\alpha),\emptyset)$,
$\T_i=(\{h_{i, 1}, \ldots,  h_{i, l_i}\}, \{\frac{\partial h_{i, 1}}{\partial y_\alpha^{(\gamma_i)}}\})$ such that $\ord(h_{i, 1}, y_\alpha)=\gamma_i\ge0$ and
\begin{equation}\label{eq-spl}
\zero(g_1, \ldots, g_l)=
 \cup_{i=1}^{N} \zero(h_{i, 1}, \ldots,  h_{i, l_i}/\frac{\partial h_{i, 1}}{\partial y_\alpha^{(\gamma_i)}})\cup\zero(\coeff(G,y_\alpha)).\end{equation}}

 1. Let $\mathbb{S}=\{\{g_1, \ldots, g_l\}\}$, $\D=\emptyset$.

 2. If $\mathbb{S}=\emptyset$, return $\D$; else let $\F\in \mathbb{S}$ and $\mathbb{S}=\mathbb{S}\setminus \{\F\}.$

 3. If $\forall f\in \F$, $\ord(f, y_\alpha)=0$, then add $(\F,\emptyset)$ to $\D$. Go to step 2.

 4. Let $f\in \F$ such that $\ord(f, y_\alpha)=t\geq 0$,
 $\deg(f, y_\alpha^{(t)})=d$. Set $\F=\F\setminus\{f\}$.\\

 5. Let  $f=\sum_{j=0}^d l_d (y_\alpha^{(t)})^d$ and
 $f_i=\frac{\partial^i f}{\partial (y_\alpha^{(t)})^i},i=1,\ldots,d$.\\

 6.
  Let $\D=\D\bigcup\{ (\{f\}\cup\F, \{f_1\}),  (\{f_1, f\}\cup\F, \{f_2\}), \ldots, (\{f_{d-1}, \ldots, f\}\cup\F, \{f_d\})\}$,  and $\mathbb{S}=\mathbb{S}\bigcup \{\F\cup\{l_0,l_1, \ldots, l_d\}\}$. Go to step 2.
\end{algorithm}
Note that the order and degree of the difference polynomials in the output are smaller than or equal to that of $g_i$ in the input.
We now give the decomposition algorithm.
\begin{algorithm}[H]\label{alg-satZ2}
 \caption{\bf --- Differential Triangular Decomposition} \smallskip
\Inp{$\{h_1, \ldots, h_k\}\in \K\{y_1, \ldots, y_n\}$.} \\
\Outp{$\{\A_q, D_q\},1=1,\ldots,N$, where $\A_q=\{\Psi_{q, 1}, \ldots \Psi_{q, l_q}\}$ satisfies the conditions in Theorem \ref{th-dm}. }

\noindent
1. Let $\T=\{\{\}, \{h_1, \ldots h_k\}, \{\}\}$,  $\mathbb{S}=\{\T\}$, $\R=\{\}$.\\
2. If $\mathbb{S}= \emptyset$,  output $\R$,  else let $\T=\{\F, \P, \N\}\in \mathbb{S}$ and $\mathbb{S}=\mathbb{S}\setminus \{\T\}$.\\
3. If $\P=\emptyset$,  add $(\F, \prod_{p\in\N} p)$ to $\R$ and go to step 2.\\
4. Let $y_\alpha^{(\gamma)} =\max_{h\in \P}\ld(h)$,
 $\widetilde{\P}=\{h\in \P\,|\,\ld(h)=y_\alpha^{(\gamma)}\}$,
 $\P=\P\setminus \widetilde{\P}$.\\
5. Let $\widetilde{\N}=\{f\in \N\,|\,\ld(f)\leq y_\alpha^{(\gamma)}\}$,  $H=\prod_{f\in \widetilde{\N}}f$.\\
6. Apply Algorithm 3 to $\{\widetilde{\P}, H,\vars(\widetilde{\P}\cup\{H\})\setminus\{y_\alpha^{(\gamma)}\}, y_\alpha^{(\gamma)}\}$, the output is $\D$.\\
7. If $\D= \emptyset $,  go to step 2,  else for $\T_1=\{\mathbb{W}=\{\psi \},  \mathbb{U}, \mathbb{V}=\{v\}\}\in \D$,  $\D=\D\setminus \{\T_1\}$,  $\mathbb{U}=\mathbb{U}\cup \P$.\\
8. If $\U=\emptyset$, add $\{\F\cup \mathbb{W}, \U, \N\cup \V\}$ to $\R$ and go to step 7. \\
9. Apply Algorithm 5 to $\{\mathbb{U}, y_\alpha\}$,  the output is $\mathbb{D}_1$.\\
10. If $\D_1=\emptyset$,  go to step 7,
 else let $C=(\Gamma, \Theta)\in \D_1$ and $\D_1=\D_1\setminus\{C\}$.\\
11. If $\Theta\neq \emptyset$,  assume $\Theta=\{\frac{\partial g}{\partial y_{\alpha}^{(l)}}\}$.
 Applying Algorithm 4 to $\{\{g\}, \mathbb{W}\cup (\Gamma\setminus\{g\}),  \mathbb{V}\}$,  the output is $\{\tilde{\mathbb{W}}, \tilde{\mathbb{V}}\}$.
 Add $\{\F, \tilde{\mathbb{W}}, \N\cup\tilde{\V}\}$ to $\mathbb{S}$. Goto step 10.\\
12. If $\Theta=\emptyset$,  let $y_\epsilon^{(r)}=\max_{h\in \Gamma}\ \ld(h)$,  $y_\beta^{(t)}=\ld(v)$.\\
13. If $y_\epsilon^{(r)}\leq y_\beta^{(t)}$ add $\{\F\cup\mathbb{W}, \Gamma, \V\cup\N \}$ to $\mathbb{S}$. Goto step 10.\\
14. Let $\textbf{x}=\{y_\gamma^{(e)}\,|\,\deg(v,y_\gamma^{(e)})>0 \hbox{ and } y_\gamma^{(e)}> y_\epsilon^{(r)}\}$ and write $v$ as a multivariate polynomial in $\textbf{x}$: $v=\Sigma l_{\theta}\textbf{x}^{\theta}$. Add
$\{\F\cup \mathbb{W}, \Gamma, \N\cup\mathbb{V} \cup \{l_\theta\}\}$ to $\mathbb{S}$ for each $\theta$. Go to step 10.\\
%
%
\end{algorithm}

We use two examples to illustrate Algorithm 6.
\begin{example}\label{ex-d1}
Note that Algorithm 6 can also be used to algebraic polynomial systems and return a radical decomposition. Let $f=xy^2$ with $x<y$.
Using Algorithm 2 to $f$,  we obtain two components $\{x\}$ and $\{y^2\}$, where the second one is not radical.
In step 6 of Algorithm 6, when applying Algorithm 3 to $\{\{f\}, \{\},y\}$,  the system $\{f=0\}$ is first split into $\{xy^2=0, 2xy\ne0)$, $\{xy^2=2xy=0, x\ne0)$, and $\{x=0\}$ and then returns $\emptyset$, $\{\{y\},\{\}, \{x\}\}$, and $\{\{\},\{x\},\{\}\}$.
Finally, we obtain the decomposition $\zero(f)=\zero(x)\cup\zero(y)$.
\end{example}

\begin{example}\label{ex-d2}
Let $f=y'^2-xy^2$,  $x<y$.
%
In step 4, we have $y_\alpha^{(\gamma)}=y'$,  $\tilde{\P}=\{f\}$.
In Step 5, $\N=\emptyset$ and $H=1$.
In Step 6, Algorithm 3 is applied to $\{\tilde{\P},H,y'\}$. $\tilde{\P}$ is first split into two components $\{y'^2-xy^2=0, 2y'\ne0\}$
and $\{y'^2-xy^2=2y'=0\}$. The output of the first component is $\{\{y'^2-4xy\}, \{\}, \{xy^2\}\}$ and the output of the second one is $\{\{y'\}, \{xy^2\}, \{\}\}$.

In Step 8,
$C_0=\{\{y'^2-xy^2\}, \{\}, \{xy^2\}\}$ will be put into $\mathbb{S}$ and eventually be added to $\R$.

In Step 9, we will handle $\{\{y'\}, \{xy^2\}, \{\}\}$.
Applying Algorithm 5 to $\U=\{xy^2\}$ and $y_\alpha=y$, the output $\D_1$ consists of $C_1=(\{xy^2\}, \{2xy\})$, $C_2=(\{xy^2,2xy\}, \{2x\})$, and $C_3=(\{2x\}, \{\})$.

$C_1$ is handled in Step 11. Algorithm 4 is applied to
$\{\{xy^2\},\{y'\},\{2xy\}\}$ and returns  $\{\{xy^2,x'y^2\},\{2xy\}\}$.
Finally, $C_4=\{\{\},\{xy^2,x'y^2\},\{2xy\}\}$ is added to $\mathbb{S}$.

$C_2$ is handled in Step 11. Algorithm 4 is applied to
$\{\{2xy\},\{y',xy^2\},\{2x\}\}$ and returns  $\{\{2xy,2x'y,xy^2\},\{2x\}\}$.
Finally, $C_5=\{\{\},\{2xy,xy^2,2xy,x'y\},\{2x\}\}$ is added to $\mathbb{S}$.

$C_3$ is handled in Steps 12 and 13. $C_6=\{\{y'\},\{2x\},\{\}\}$ is added to $\mathbb{S}$.

For $C_4$, in Step 6, Algorithm 3 is applied to
$\{\{xy^2,x'y^2\},1,\{2xy\}\}$ and returns the empty set.
We omit the computing procedures for $C_5$ and $C_6$.
The algorithm give the decomposition $\zero(f)=\zero(\sat(f))\cup\zero(y',x)\cup\zero(y)$.
\end{example}

Now we prove Theorem \ref{th-dm} with the following lemmas.

\begin{lemma}
Algorithm 6 terminates,  $\zero(h_1, \ldots,  h_k)=\cup_q \zero(\A_q/D_q)$, and
$I_{\Psi_{q, i}}$,  $S_{\Psi_{q, i}}\neq 0$ on any element of $\zero(\A_q/D_q)$.
\end{lemma}
\proof
The algorithm has three loops, starting at Step 2, Step 7, and Step 10, respectively. We need only to show that the loop starting at Step 2 will terminate. Let $\{\F_1,\P_1,\N_1\}$ be a component added to $\mathbb{S}$ in this loop and $y_{c}^{\delta} = \max_{p\in \P_1} \ld(p)$.
Then, we have either $y_{c}^{\delta}< y_{\alpha}^{\gamma}$ which means that the algorithm terminates.

$\zero(h_1, \ldots, h_k)=\bigcup\limits_q \zero(\Psi_{q, 1}, \ldots \Psi_{q, l_q}/D_q)$ can be proved similar to Lemma \ref{lm-2}. In the proof, we also need the equalities in Lemma \ref{differential gcd},  Lemma \ref{prem},  and \bref{eq-spl}.

We now show that $\A_q$ is a triangular set.
It suffices to show that for any $\{\F,\P,\N\}\in \mathbb{S}$, $\max_{p\in \P} \cls(p) < \max_{q\in \F} \cls(q)$. New polynomials are added to $\F$ in Steps 8, 13, and 14. In Step 8, since $\U=\emptyset$, this is indeed the case.
In Steps 13 and 14, we have $\Theta=\emptyset$ which means that $y_\alpha$ and its derivatives do not appear in $\Gamma$. Hence $\max_{p\in \Gamma} \cls(p) < \alpha$ and $\A_q$ is a triangular set for any $q$.

Finally,  if $(\Psi_1, \ldots, \Psi_t/M)$ is one component of the output,  then according to the algorithm it comes from a procedure like \bref{eq-pr1} and \bref{eq-pr2}.
In the algebraic case, from one step to the next step in \bref{eq-pr1}, Algorithm 1 is used one time.
In the differential case, from one step to the next step in \bref{eq-pr1}, Algorithm 3 is used many times. For instance, the procedure to obtain $\Psi_1$ is as follows:
 \begin{eqnarray}
\zero(f_{0, 1}, \ldots , f_{0, k^{(0)}}/M_0)
 &\rightarrow& \zero(\psi_1, h_{1, 1}, \ldots , h_{1, t^{(1)}}/M_1)\nonumber\\
 &\rightarrow&\zero(g_{1, 0}, \ldots, g_{1, l^{(1)}}/S_1M_1)\nonumber\\
 &\rightarrow& \zero(\psi_2, h_{2, 1}, \ldots , h_{2, t^{(2)}}/M_2S_1M_1)\nonumber\\
 &\rightarrow&\ldots\\
 &\rightarrow&\zero(g_{s, 0},\ldots,g_{s,l^{(s)}}/M_{s+1}\cdots S_1M_1)\nonumber\\
 &\rightarrow&\zero(\psi_{s+1}, h_{s+1, 1}, \ldots , h_{s+1, t^{(s+1)}}/S_{s+1}M_{s+1}\cdots S_1M_1)\nonumber
\end{eqnarray}
where $\Psi_{1} = \psi_{s+1}$ and $\{h_{s+1, 1}, \ldots , h_{s+1, t^{(s+1)}}\} = \{f_{1,1},\ldots,f_{1,k^{(1)}}\}$ in \bref{eq-pr1}.
$(\psi_1, h_{1, 1}, \ldots,$ $h_{1, t^{(1)}}/M_1)$ is a component of $(f_{0, 1}, \ldots , f_{0, k^{(0)}}/M_0)$ after using Algorithm 3, so $I_{\psi_1}|M_1$ and
$S_{\psi_1}\neq 0$ on any element of $\zero(\psi_1, h_{1, 1}, \ldots , h_{1, k^{(1)}}/M_1)$ by Lemma \ref{differential gcd}.
$(g_{1, 0}, \ldots, g_{1, l^{(1)}}/S_1M_1)$ is a component obtained from
$(\psi_1, h_{1, 1}, \ldots , h_{1, k^{(1)}}/M_1)$ by Algorithms 5 and 6 in Steps 9 and 11. So we have
$\zero(g_{1, 0}, \ldots, g_{1, l^{(1)}}/S_1M_1)\subset \zero(\psi_1, h_{1, 1}, \ldots , h_{1, k^{(1)}}/M_1)$.
The procedure is repeated until $\cls(\psi_{s+1}) > \cls(g_{s+1, j})$ for all $j$ and $\Psi_1$ is obtained. Then, we have
 \begin{eqnarray}
 \zero(f_{0, 1}, \ldots , f_{0, k^{(0)}}/M_0)&\supseteq&
 \zero(\psi_1, h_{1, 1}, \ldots , h_{1, t^{(1)}}/M_1)\nonumber\\
&\supseteq&\zero(g_{1, 0}, \ldots, g_{1, l^{(1)}}/S_1M_1)\nonumber\\
&\supseteq&\ldots\\
&\supseteq&\zero(g_{s, 0},\ldots,g_{s,l^{(s)}}/S_{s}\cdots S_1M_1)\nonumber\\
&\supseteq&\zero(\psi_{s+1}, h_{s+1, 1}, \ldots , h_{s+1, t^{(s+1)}}/M_{s+1}\cdots S_1M_1)\nonumber
\end{eqnarray}
By Lemma \ref{differential gcd}, $I_{\Psi_1}|M_{s+1}$ and
$S_{\Psi_1}\neq 0$ on any element of $\zero(\psi_{s+1}, h_{s+1, 1}, \ldots , h_{s+1, t^{(s+1)}}/M_{s+1}$ $\cdots S_1M_1)$.
The lemma is proved.\qedd

\begin{lemma}In Algorithm 6,  $\A_q :=\Psi_{q, 1}, \ldots \Psi_{q, l_q}$ are saturated triangular sets and\\
$\zero(\dsat(\A_q))=\overline{\zero(\Psi_{q, 1},  \ldots \Psi_{q, l_q}/D_q)}$.
\end{lemma}
\proof Using Lemmas \ref{differential regular} and \ref{differential zariski} instead of Lemmas \ref{irreducible condition} and \ref{zariski  closure lemma},  the proof of this lemma is the same with that of
Lemma \ref{lm-3}.\qedd

The following lemma gives the complexity part of Theorem \ref{th-dm}.
\begin{lemma} In Algorithm 6,  the degree of $\Psi_{q, i}$ is less than $d^{c^{2^nR}}$,  the order of $\Psi_{q, i}$ is less than $2^nR$,  $N<k^{2^nR}d^{c^{2^nR}Rn}$.
The running time of this algorithm can be bounded by a polynomial in $k^{2^nR}d^{c^{2^nR}Rn}$.
\end{lemma}
\proof
%
For given differential polynomials $h_1, \ldots, h_k\in\K\{y_1, \ldots, y_n\}$,  since $\ord(h_i, y_j)<R$ for $1\leq i\leq k, 1\leq j\leq n$,  there are at most $Rn$ variables
when applying Algorithm 3.
Consider the most complicated case where $y_n^{(R-1)}$ is the maximal leader.
After applying Lemma \ref{differential gcd} to $h_1, \ldots, h_k$,  there are at most $kd^{cRn}$ components,  each component has at most $kd^{cRn}$ differential polynomials, and the degree of each polynomial is no more than $d^c$.
After applying Algorithm 5,  each component will be split into at most $d^{cR}$ components,  and each component has at most
$kd^{cRn+cR}$ polynomials.
After applying Lemma \ref{prem},  the number of the components and the number of polynomials in each component do not change,  and the degrees of the polynomials are less than $d^{2c}$.
The most complicated case occurs when the order of $y_n$ decreases by one and the maximal leader becomes $y_n^{(R-2)}$. Continuing this procedure until the main variable becomes $y_{n-1}$. There are
no more than $k^Rd^{c^RRn}$ components in total,  each component has no more than $kd^{c^RRn}$ polynomials,  the degree of these polynomials are less than $d^{c^R}$, and the order of these polynomials are less than $2R$. Repeating this procedure to $y_1, \ldots, y_{n-1}$,  finally we obtain no more than $k^{2^nR}d^{c^{2^nR}Rn}$ components,  the degree of the polynomials are bounded by $d^{c^{2^nR}}$,  and the order of these polynomials are less than $2^nR$.\qedd

\section{Summary}
Two triangular decomposition algorithms are given in this paper.
For a set of polynomials $F=\{f_1,\ldots,f_s\}$ in $\K[x_1,\ldots,x_n]$, we can compute regular triangular sets $\T_1,\ldots,\T_r$ such that
$\zero(F)=\cup_i\zero(\sat(\T_i))$ which gives an unmixed decomposition for the solution set of $F=0$. We also show that the complexity of the algorithm is double exponential in $n$.
For a set of ordinary differential polynomials $F=\{f_1,\ldots,f_s\}$ in $\K\{y_1,\ldots,y_n\}$, we can give a similar decomposition
$\zero(F)=\cup_i\zero(\dsat(\T_i))$, where $\dsat(\T_i)$ are radical differential ideals and the complexity is triple exponential. This seems to be the first triangular decomposition algorithm for differential polynomial systems with elementary computation complexity.

\end{document}